\documentclass[twocolumn,twocolappendix]{aastex631}

\shorttitle{SN 2023tsz}
\shortauthors{Vasylyev et al.}

\newcommand{\LCO}{\affiliation{Las Cumbres Observatory, 6740 Cortona Drive, Suite 102, Goleta, CA 93117-5575, USA}}
\newcommand{\UCSB}{\affiliation{Department of Physics, University of California, Santa Barbara, CA 93106-9530, USA}}

\newcommand{\UCD}{\affiliation{Department of Physics and Astronomy, University of California, Davis, 1 Shields Avenue, Davis, CA 95616-5270, USA}}

\newcommand{\UCB}{\affiliation{Department of Astronomy, University of California, Berkeley, CA 94720-3411, USA}}

\newcommand{\STScI}{\affiliation{Space Telescope Science Institute, 3700 San Martin Drive, Baltimore, MD 21218-2410, USA}}
\newcommand{\UT}{\affiliation{Department of Astronomy, University of Texas at Austin, 1 University Station C1400, Austin, TX 78712-0259, USA}}

\newcommand{\IPAC}{\affiliation{Spitzer Science Center, California Institute of Technology, Pasadena, CA 91125, USA}}

\newcommand{\CfA}{\affiliation{Center for Astrophysics \textbar{} Harvard \& Smithsonian, 60 Garden Street, Cambridge, MA 02138-1516, USA}}
\newcommand{\UA}{\affiliation{Steward Observatory, University of Arizona, 933 North Cherry Avenue, Tucson, AZ 85721-0065, USA}}

\newcommand{\JHU}{\affiliation{William H. Miller III Department of Physics and Astronomy, Johns Hopkins University, Baltimore, MD 21218, USA}}
\newcommand{\ARCO}{\affiliation{Astrophysics Research Center of the Open University (ARCO), Department of Natural Sciences, Ra’anana 4353701, Israel}}

\newcommand{\GeminiNorth}{\affiliation{Gemini Observatory, 670 North A`ohoku Place, Hilo, HI 96720-2700, USA}}

\newcommand{\Rutgers}{\affiliation{Department of Physics and Astronomy, Rutgers, the State University of New Jersey,\\136 Frelinghuysen Road, Piscataway, NJ 08854-8019, USA}}

\newcommand{\TAMU}{\affiliation{Department of Physics and Astronomy, Texas A\&M University, 4242 TAMU, College Station, TX 77843, USA}}

\newcommand{\Szeged}{\affiliation{Department of Experimental Physics, University of Szeged, D\'om t\'er 9, Szeged, 6720, Hungary}}
\newcommand{\IAP}{\affiliation{Institut d'Astrophysique de Paris, CNRS--Sorbonne Universit\'e, 98 bis boulevard Arago, F-75014 Paris, France}}
\newcommand{\UCSD}{\affiliation{Department of Astronomy \& Astrophysics, University of California, San Diego, 9500 Gilman Drive, MC 0424, La Jolla, CA 92093-0424, USA}}
\newcommand{\Monash}{\affiliation{School of Physics and Astronomy, Monash University, Clayton, Victoria 3800, Australia}}
\newcommand{\OzGrav}{\affiliation{OzGrav: The ARC Centre of Excellence for Gravitational Wave Discovery, Clayton, Victoria 3800, Australia}}

\begin{document}

\title{Ultraviolet to Infrared Spectroscopy of the Type Ibn SN 2023tsz Suggests a Lower-mass Progenitor}

\correspondingauthor{Sergiy Vasylyev}
\email{svasylyev@ucsd.edu}


\author[0000-0002-4951-8762]{Sergiy S. Vasylyev}
\UCSD

\author[0000-0002-0832-2974]{Griffin Hosseinzadeh}
\UCSD

\author[0000-0003-0599-8407]{Luc Dessart}
\IAP
\affiliation{French-Chilean Laboratory for Astronomy, IRL 3386, CNRS, Instituto de Astrofísica, Pontificia Universidad Católica de Chile, Casilla 306, Santiago, Chile}

\author[0000-0003-2238-1572]{Ori Fox}
\STScI

\author[0000-0003-4102-380X]{David J.\ Sand}
\UA

\author[0000-0003-3460-0103]{Alexei V. Filippenko}
\UCB 

\author[0000-0001-5233-6989]{Qinan Wang}
\affiliation{Kavli Institute for Astrophysics and Space Research, Massachusetts Institute of Technology, Cambridge, MA 02139, USA}

\author[0000-0003-4253-656X]{D. Andrew Howell}
\UCSB
\LCO

\author[0000-0003-0123-0062]{Jennifer E.\ Andrews}
\GeminiNorth

\author[0000-0002-1895-6639]{Moira Andrews}
\LCO\UCSB

\author{Pallas Beddow}
\UCB

\author{K. Azalee Bostroem}
\UA

\author[0000-0001-5955-2502]{Thomas G. Brink}
\UCB

\author[0000-0001-6272-5507]{Peter J. Brown}
\TAMU

\author{Asia deGraw}
\UCB

\author[0000-0002-7937-6371]{Yize Dong}
\CfA


\author[0000-0003-4914-5625]{Joseph Farah}
\LCO\UCSB

\author[0000-0003-2824-3875]{Thomas R. Geballe}\affiliation{Gemini Observatory/NSF's National Optical-Infrared Astronomy Research Laboratory, 670 N. Aohoku Place, Hilo, HI, 96720, USA}

\author{Sebastian Gomez}
\UT

\author[0000-0003-2744-4755]{Emily T. Hoang}
\UCD

\author[0000-0002-9454-1742]{Brian Hsu}
\UA

\author[0000-0001-8738-6011]{Saurabh~W.~Jha}
\Rutgers

\author{Patrick Kelly}

\author[0000-0001-5807-7893]{Curtis McCully}
\LCO
\author[0009-0008-9693-4348]{Darshana Mehta}
\UCD

\author[0000-0001-9570-0584]{Megan Newsome}
\UT

\author[0000-0003-3656-5268]{Yuan Q. Ni}
\UCSB
\LCO

\author[0000-0001-7488-4337]{Seong Hyun Park}\affiliation{Department of Physics and Astronomy, Seoul National University, Gwanak-ro 1, Gwanak-gu, Seoul, 08826, South Korea}

\author[0000-0002-0744-0047]{Jeniveve Pearson}
\UA

\author{Neil Pichay}
\UCB

\author{Justin Pierel}
\STScI

\author[0000-0002-7352-7845]{Aravind P.\ Ravi}
\UCD

\author[0000-0002-7015-3446]{Nicolas E.\ Meza Retamal}
\UCD
\author{Melissa Shahbandeh}
\STScI

\author[0000-0002-4022-1874]{Manisha Shrestha}
\Monash\OzGrav

\author{Nathan Smith}
\UA

\author[0000-0001-8073-8731]{Bhagya M.\ Subrayan}
\UA
\author[0000-0003-4610-1117]{Tam\'as Szalai}
\Szeged
\author[0000-0001-8818-0795]{Stefano Valenti}
\UCD
\author{Schuyler D. Van Dyk} 
\IPAC
\author[0000-0003-3643-839X]{Jeonghee Rho}\affil{SETI Institute, 189 Bernardo Ave., Ste. 200, Mountain View, CA 94043, USA}

\author[0009-0006-7296-728X]{Kathryn Wynn}
\LCO \UCSB 

\author[0000-0002-6535-8500]{Yi Yang}
\affil{Department of Astronomy, University of California, Berkeley, CA 94720-3411, USA}
\affil{Physics Department, Tsinghua University, Beijing, 100084, China}

\author[0000-0002-0632-8897]{Yossef Zenati}  \ARCO \JHU

\author[0000-0002-2636-6508]{WeiKang Zheng}
\UCB

\begin{abstract}
Type Ibn supernovae are stripped-envelope explosions whose spectra indicate interaction with dense, helium-rich and hydrogen-poor circumstellar material (CSM), making them important probes of late-stage mass loss and progenitor stripping.
We present extensive ultraviolet-to-near-infrared spectrophotometry of the Type Ibn SN\,2023tsz, including two epochs of HST/STIS ultraviolet (UV) spectroscopy and ground-based optical and near-infrared follow-up observations. The spectra are dominated by intermediate-width emission lines at all phases after maximum light, suggesting that much of the luminosity originates in a cold dense shell (CDS) formed by interaction between the ejecta and CSM. We compare the observations to one-dimensional non-local-thermodynamic-equilibrium radiative-transfer models of a 4\,M$_{\odot}$ (at initial helium burning) He-star explosion, which reproduce the strong optical and near-infrared \ion{He}{1} lines and require an added X-ray irradiation field to match the highly ionized UV features.
The spectra are best reproduced by models with an X-ray irradiation power of $L_\mathrm{X} \approx 10^8\,{\rm L}_{\odot}$, with the preferred models favoring CDS radii of order $(1.5$--$2)\times10^{15}$\,cm, velocities of $\sim5\times10^7$\,cm\,s$^{-1}$, and interaction powers of a few $\times10^{42}$\,erg\,s$^{-1}$. In the optical, the preferred models shift from higher interaction power and smaller radii at early times to lower power and larger radii at later times. These results add to the growing evidence that at least some SNe Ibn arise from lower-mass helium stars whose final evolution is shaped by binary interaction.

\end{abstract}

\keywords{Circumstellar matter (241), Core-collapse supernovae (304), Stellar mass loss (1613), Supernovae (1668)}

\section{Introduction} \label{sec:intro}

Type Ibn supernovae (SNe Ibn) represent a rare subset of stripped-envelope supernovae (SESNe) in which the ejecta interact with dense, helium-rich and hydrogen-poor circumstellar material (CSM). Their defining observational signature is the presence of relatively narrow \ion{He}{1} emission lines (full width at half-maximum intensity [FWHM] $\approx 1000$--2000 km s$^{-1}$; \citealt{hosseinzadeh_type_2017,dong_spectral_2025}), analogous to the intermediate-width hydrogen lines seen in spectra of Type IIn SNe \citep{filippenko_optical_1997,smith_massive_2011,taddia_carnegie_2013}. The prototype event for SNe Ibn was SN~1999cq \citep{matheson_helium_2000}, which showed intermediate-width \ion{He}{1} features without any detectable H$\alpha$, indicating an explosion within a He-dominated environment.

SN\,2006jc cemented SNe Ibn as a distinct class. It showed relatively narrow \ion{He}{1} lines superposed on a Type Ib-like continuum and, remarkably, was preceded by a luminous outburst observed two years prior to explosion \citep{foley_sn_2007,pastorello_giant_2007,pastorello_massive_2008}.
The precursor eruption reached $M_R = -14$ mag and resembled the giant outbursts of luminous blue variable stars, despite producing nearly hydrogen-free CSM. This provided the first direct link between SNe~Ibn and episodic mass loss shortly before core collapse. Subsequent SNe Ibn have corroborated this picture: precursor outbursts have been detected in SN~2019uo \citep{strotjohann_bright_2021} and most recently SN\,2023fyq, which exhibited multiyear variability with a dramatic brightening in the final $\sim$100 days prior to explosion \citep{brennan_spectroscopic_2024,dong_sn2023fyq_2024-1}. These observations confirm that intense mass-loss episodes in the final years are a common feature of at least some SN Ibn progenitors.

Early-time spectra of some SNe~Ibn have exhibited fleeting ``flash'' features analogous to those seen in some interacting SNe II (e.g., SN\,1998S, \citealt{leonard_evidence_2000,shivvers_early_2015}; SN\,2013fs, \citealt{yaron_confined_2017}). For instance, SN\,2010al showed extremely narrow, high-ionization lines such as He~\textsc{ii}, C~\textsc{iii}, and N~\textsc{iii} that disappeared within days as the ejecta overtook the dense inner CSM \citep{pastorello_massive_2015}. Similar features have been reported in SN\,2019uo \citep{gangopadhyay_flash_2020}, SN\,2019wep \citep{gangopadhyay_evolution_2022}, and SN\,2023emq \citep{pursiainen_sn_2023}. These lines imply the presence of a compact, dense CSM shell ejected just months to years before core collapse of their progenitors.

Over the past two decades, systematic studies have expanded the sample of SNe Ibn and revealed common trends in their photometric and spectroscopic evolution \citep{pastorello_massive_2016,moriya_circumstellar_2016,hosseinzadeh_type_2017,farias_characterization_2026}. While SNe~Ibn generally exhibit rapid, homogeneous light curves, their spectral properties show diversity \citep{hosseinzadeh_type_2017,dong_spectral_2025}, with some displaying narrow P-Cygni absorption components and others showing purely emission-line profiles. SNe Ibn also have varying widths in their narrow-line profiles. This diversity likely reflects differences in the density and geometry of the CSM, and therefore differences in mass-loss history.

The environments and host galaxies of SNe~Ibn provide further clues to their origins. Several  well-studied SNe~Ibn have been found outside of star-forming regions, inconsistent with massive, young progenitors. PS1-12sk, for example, occurred in the outskirts of an elliptical galaxy devoid of recent star formation \citep{sanders_ps1-12sk_2013,hosseinzadeh_type_2019}. Studies of local stellar populations around SN\,2006jc and SN\,2015G indicate older environments inconsistent with very massive stars, pointing toward a lower-mass binary origin in some cases \citep{pastorello_massive_2015b,sun_origins_2020}. Direct evidence for a binary channel comes from the detection of a surviving companion star to SN\,2006jc in late-time \textit{HST} imaging \citep{maund_possible_2016,sun_origins_2020}.

Connections between SNe Ibn and other rare transient classes have also been explored. Some properties overlap with those of fast blue optical transients, such as AT~2018cow \citep{fox_signatures_2019}, suggesting a continuum of interaction-powered phenomena \citep{ho_search_2023}. Moreover, several events with even more extreme stripping have been classified as Type Icn SNe, exhibiting narrow C/O emission instead of helium (SN~2010mb, \citealt{ben-ami_sn_2014}; SN~2019hgp, \citealt{gal-yam_wc_2022}; SNe 2019jc and 2021ckj, \citealt{pellegrino_diverse_2022}; SN~2021csp, \citealt{fraser_sn_2021}, \citealt{perley_type_2022}; SN~2022ann, \citealt{davis_sn_2023}; and SN~2021ocs, \citealt{kuncarayakti_late-time_2022}), or even Type~Ien, interacting with Si/S-rich material (although the dominant species remains O; SN~2021yfj, \citealt{schulze_extremely_2025}).

Despite the insights gained from optical data, ultraviolet (UV) observations of SNe Ibn remain rare and provide unique diagnostic power. UV spectra directly probe the ionization state and chemical composition of the outermost material, including the structure of the CSM. 
The UV also probes strong resonance lines of multiple ionization
stages of key elements like C (e.g., \ion{C}{2}, \ion{C}{3}, and \ion{C}{4}), and thus helps
identify the influence of X-ray photoionization.  Only a handful of SNe Ibn have been observed in the UV: SN~2006jc with the Neil Gehrels Swift Observatory \citep{immler_swift_2008,bufano_ultraviolet_2009}, and SN~2010al \citep{kirshner_hst_2010,pastorello_massive_2015}, SN~2015G \citep{filippenko_early-time_2014,shivvers_nearby_2017}, and SN~2020nxt \citep{fox_uv_2019,wang_low-mass_2024} with the Hubble Space Telescope (HST).


Recent theoretical work has explored a variety of progenitor channels and mass-loss mechanisms that could give rise to the dense, He-rich CSM inferred for SNe~Ibn \citep{ercolino_mass-transferring_2025}. Light-curve modeling of ejecta interacting with a compact, dense Wolf-Rayet (WR) wind has shown that such configurations can reproduce the fast, luminous evolution of some SNe~Ibn \citep{maeda_properties_2022}. More exotic scenarios have also been proposed, including pulsational pair-instability SNe \citep{woosley_pulsational_2017,renzo_predictions_2020} and mergers involving WR stars and compact objects such as black holes or neutron stars \citep{metzger_luminous_2022}. 
Using radiation hydrodynamics and a $4~\mathrm{M}_\sun$ He-star progenitor model,
\cite{dessart_helium_2022} explored the diversity of light curve properties resulting from ejecta
interaction with CSM originating from a dense, long-lived
wind or an explosively ejected shell.

Binary evolution provides another promising pathway to an SN~Ibn. Calculations have shown that mass transfer and common-envelope evolution can produce low-mass helium stars embedded in residual CSM or a circumbinary disk \citep{laplace_expansion_2020,laplace_different_2021,tuna_long-term_2023,ercolino_mass-transferring_2025}, while  modeling of explosions of such low-mass He-star progenitors interacting with dense, He-rich CSM reproduces many of the observed properties of SNe~Ibn \citep{dessart_helium_2022,wang_low-mass_2024}. In these relatively low-mass systems, late-stage instabilities may play a crucial role in shaping the CSM: gravity waves excited by vigorous convection in advanced burning stages can transport energy into the envelope, inflating it and reducing its binding energy without necessarily producing bright, long-lived pre-SN outbursts \citep{fuller_pre-supernova_2018,wu_diversity_2021,wu_extreme_2022,wu_wave-driven_2022}. The resulting extended, weakly bound envelope is more susceptible to stripping or reshaping by binary interaction, naturally producing compact, dense, helium-rich CSM while remaining consistent with the limited pre-explosion variability seen in many progenitor systems.

\begin{figure*}
    \centering
    \includegraphics[width=0.7\textwidth]{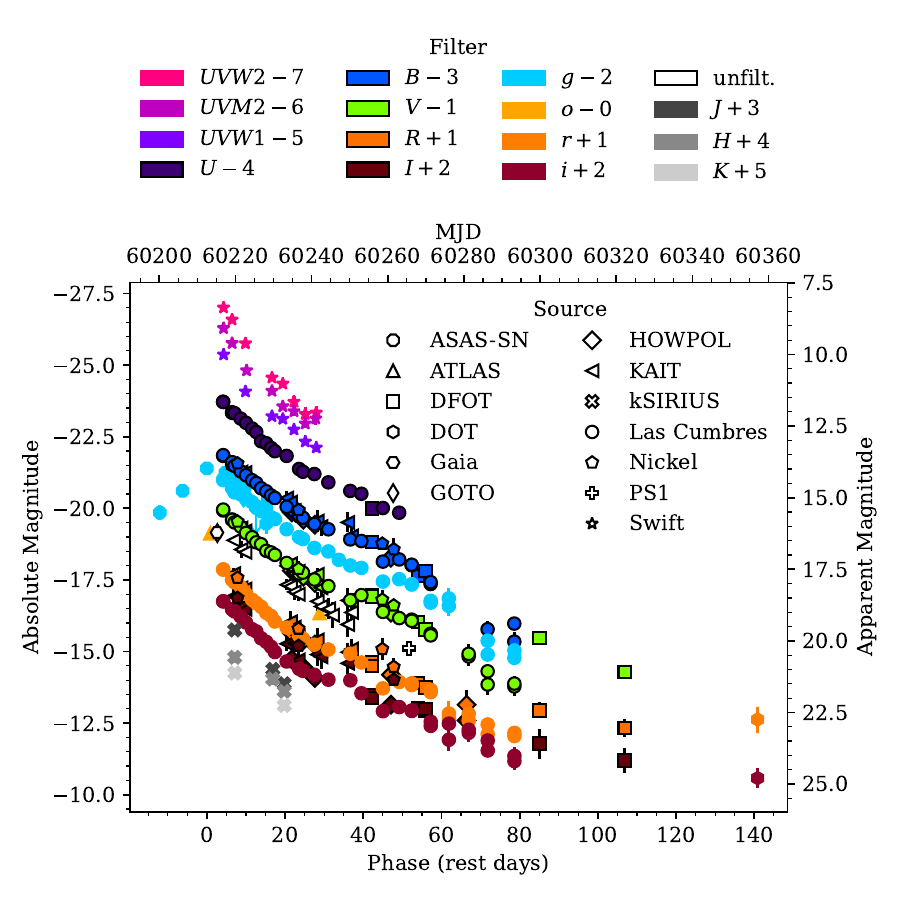}
    \caption{Summary of collected photometry for SN\,2023tsz. The phase is relative to the time of maximum brightness. Unfiltered photometry includes data from Gaia $G$, GOTO $L$, KAIT ``Clear,'' and PS1 $w$, and is not offset.}
    \label{fig:phot}
\end{figure*} 

Here we present UV, optical, and near-infrared (NIR) spectroscopy of the Type~Ibn SN\,2023tsz, which exploded in a low-luminosity but nearby dwarf galaxy. This event was previously discussed by \cite{warwick_sn_2025}, who focused mainly on the low metallicity and low star-formation rate of the host galaxy. This environment poses a problem for progenitor scenarios involving single very massive WR stars, which have relatively short delay times between formation and explosion and require high metallicity for wind-driven stripping of their hydrogen-rich envelopes. Instead, we use spectral modeling to argue that SN~2023tsz comes from a low-mass helium star stripped through binary interaction, which is fully compatible with its observed host-galaxy environment. This has implications for the greater class of SNe~Ibn and how they fit into the zoo of stellar explosions.
\begin{figure*}
    \centering
    \includegraphics[width=\textwidth]{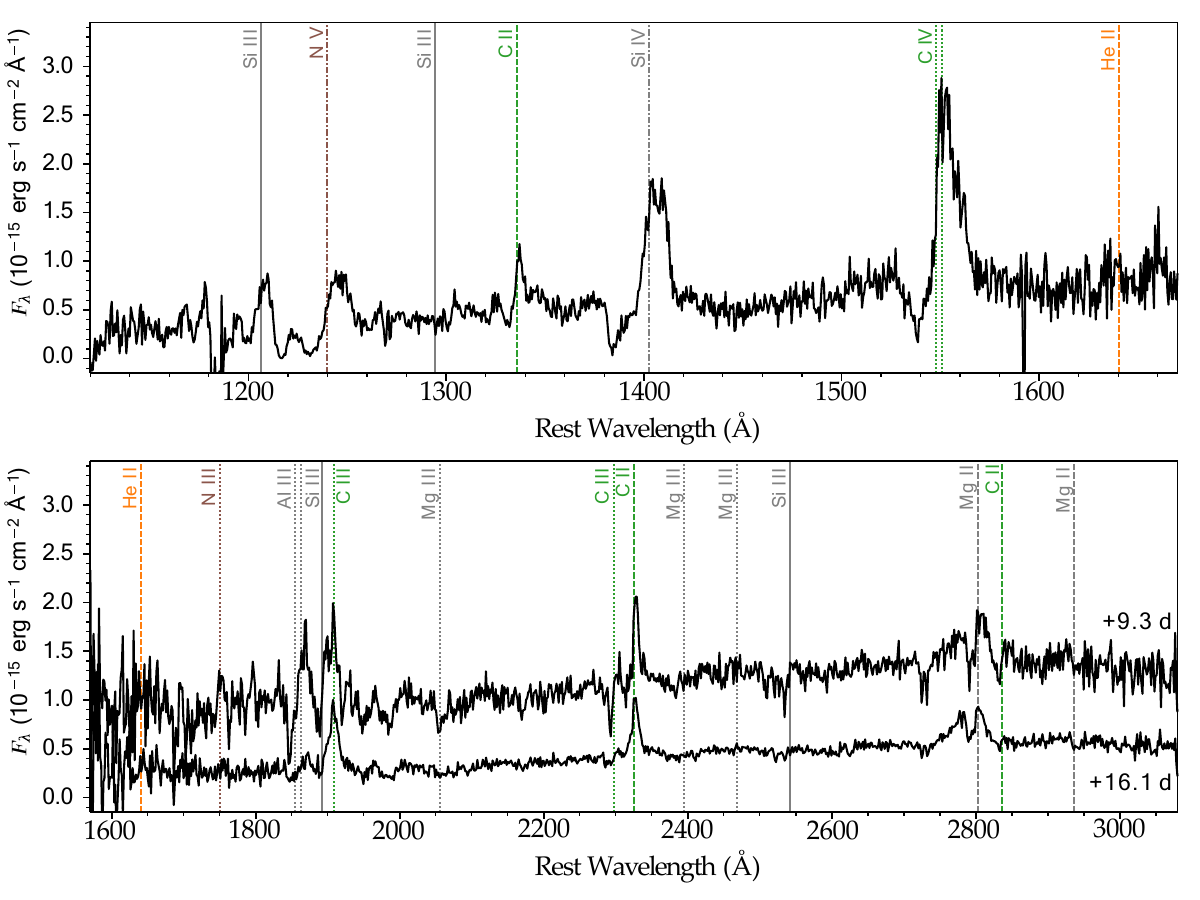}
    \caption{HST/STIS  total flux density spectra of SN\,2023tsz in the far-UV (top panel) and near-UV (bottom panel) regions. In the bottom panel, the top (bottom) near-UV spectrum 
    corresponds to +9.3\,d (+16.1\,d) after maximum light.}
    \label{fig:uvspec}
\end{figure*}
In Section \ref{sec:obs}, we present the UV to NIR observations obtained for SN\,2023tsz with ground- and space-based facilities. Section \ref{sec:modeling} describes the modeling approach used to generate synthetic spectra to compare to observations. We discuss in Section \ref{sec:results} the results of the comparison to the model grid and their implications for the progenitor and environment of SN\,2023tsz. In Section \ref{sec:conclusions}, we summarize the main takeaways of this work.
\section{Observations} \label{sec:obs}

SN~2023tsz (J2000 coordinates $\alpha=08\textsuperscript{h}37\textsuperscript{m}29\fs{}530$, $\delta=-00\degr{}02'35\farcs{}74$) was reported and classified by the Gravitational-wave Optical Transient Observer (GOTO) Collaboration \citep{steeghs_gravitational-wave_2022}. \cite{godson_goto_2023} discovered it on UT 2023 September 28 (MJD 60215.23890) with the GOTO-2 at Roque de los Muchachos Observatory (Canary Islands, Spain) at a wide-band optical brightness of $L = 16.36 \pm 0.02$~mag. The SN was just emerging from Sun constraint at the time of discovery, so there are no nondetections from any time-domain surveys in the months beforehand. \cite{pursiainen_goto_2023} classified it as a Type~Ibn SN using a spectrum taken the next day (2023 September 29; MJD 60216.232) with the the Alhambra Faint Object Spectrograph and Camera (ALFOSC) on the Nordic Optical Telescope (NOT; Canary Islands, Spain).

We obtained follow-up imaging with
the Sinistro cameras on Las Cumbres Observatory's network of 1~m telescopes \citep{brown_cumbres_2013} as part of the Global Supernova Project \citep{howell_global_2017},
the Katzman Automatic Imaging Telescope (KAIT; \citealt{filippenko_lick_2001}) and the CCD camera \citep{stone_ccd_1990} on the 1\,m Nickel telescope at Lick Observatory (CA, USA),
the Hiroshima One-shot Wide-field Polarimeter (HOWPOL; \citealt{kawabata_wide-field_2008}) 
on the 1.5~m Kanata Telescope at Higashi-Hiroshima Observatory (Hiroshima, Japan),
kSIRIUS on the 1~m optical-infrared telescope at Iriki Observatory (Kagoshima, Japan),
the 1.3~m Devasthal Fast Optical Telescope (DFOT; \citealt{joshi_aries_2022}) and the 3.6~m Devasthal Optical Telescope (DOT; \citealt{omar_optical_2019}) at Aryabhatta Research Institute of Observational Sciences' facility in Devasthal, India,
and the Ultraviolet/Optical Telescope \citep{roming_swift_2005} on the Neil Gehrels Swift Observatory \citep{gehrels_swift_2004}. See Appendix~\ref{app:phot} for details of the photometry reduction.

We performed forced photometry at the position of SN~2023tsz from the All-sky Automated Survey for SNe (ASAS-SN; \citealt{shappee_man_2014}) using Sky Patrol Version 2.0 \citep{hart_asas-sn_2023}. This photometry includes three detections prior to discovery showing a rise to peak starting on 2023 September 13 (MJD 60200.14162) at $g = 17.66 \pm 0.27$~mag ($4\sigma$ significance). Therefore, the SN was already at least two weeks old at discovery. We include discovery photometry reported to the Transient Name Server by the Asteroid Terrestrial-impact Last Alert System (ATLAS; \citealt{tonry_atlas_2018a}), Gaia Photometric Science Alerts, GOTO, and the Panoramic Survey Telescope and Rapid Response System 1 (PS1; \citealt{chambers_pan-starrs1_2016}) in our analysis. Finally, we obtained one additional detection from the ATLAS forced photometry server \citep{shingles_release_2021}. All the photometry we analyze is plotted in Figure~\ref{fig:phot}.

We obtained optical spectra of SN~2023tsz with
FLOYDS on Las Cumbres Observatory's Faulkes Telescope North (FTN; \citealt{brown_cumbres_2013}) on Haleakal\=a (HI, USA),
Binospec \citep{fabricant_binospec_2019} and Blue Channel spectrograph \citep{angel_optical_1979} on the MMT on Mt.~Hopkins (AZ, USA),
the Goodman High-throughput Spectrograph \citep{clemens_goodman_2004} on the Southern Astrophysical Research Telescope (SOAR) on Cerro Pach\'on (Chile),
the Kast double spectrograph \citep{miller_kast_1994} on the C.~Donald Shane Telescope at Lick Observatory (CA, USA),
the Low-resolution Imaging Spectrometer (LRIS; \citealt{oke_keck_1995}) on the Keck~I Telescope on Maunakea (HI, USA), 
the Himalayan Faint Object Spectrograph Camera (HFOSC) on the Himalayan Chandra Telescope (HCT) at the Indian Astronomical Observatory (Hanle, India),
the ARIES-Devasthal Faint Object Spectrograph and Camera (ADFOSC; \citealt{omar_optical_2019}) on the Devasthal Optical Telescope (DOT) at Devasthal Observatory (Nainital, India),
and the Multi-object Double Spectrographs (MODS; \citealt{pogge_multi-object_2006}) on the Large Binocular Telescope (LBT) on Mt.~Graham (AZ, USA). We also include the classification spectrum of \cite{pursiainen_goto_2023} in our analysis. We obtained an NIR spectrum using the Gemini Near-infrared Spectrograph (GNIRS; \citealt{elias_design_2006,elias_performance_2006}) on the Gemini North telescope on Maunakea (HI, USA). See Appendix~\ref{app:spec} for details of the ground-based spectroscopy reduction.

Finally, we obtained two epochs of UV spectroscopy using the Space Telescope Imaging Spectrograph (STIS; \citealt{woodgate_space_1998}) multianode microchannel arrays (MAMAs; \citealt{timothy_review_2016}) on the Hubble Space Telescope as part of program GO-17195 \citep{fox_uv_2022}. In the first epoch, taken 21.3 days after first detection, we used both the near-UV (NUV) MAMA with the G230L grism and the far-UV (FUV) MAMA with the G140L grism, covering 115--310~nm. In the second epoch, taken 28.1 days after first detection, we used only the NUV-MAMA with G230L, covering 180--310~nm, owing to the rapid decrease in UV brightness expected for SNe~Ibn. The reduced UV spectra are available in the Mikulski Archive for Space Telescopes (DOI: \dataset[10.17909/ym0y-s612]{\doi{10.17909/ym0y-s612}}) and are shown in Figure \ref{fig:uvspec}.

All our spectra are logged in Table~\ref{tab:spec}, are available in machine-readable form as data behind Figures~\ref{fig:uvspec}, 
\ref{fig:spec_groups}, and \ref{fig:best_fit_nir_new}, and are archived in the Weizmann Interactive Supernova Data Repository \citep{yaron_wiserep_2012}.
\begin{deluxetable}{ccccc}
\tablecaption{Log of Spectroscopic Observations \label{tab:spec}}
\tablehead{\colhead{MJD} & \colhead{Telescope} & \colhead{Instrument} & \colhead{Phase} & \colhead{Notes}}
\startdata
60216.232 & NOT & ALFOSC & +3.6 & \tablenotemark{a} \\
60220.605 & FTN & FLOYDS & +7.9 \\
60221.492 & MMT & Blue Channel & +8.8 & \tablenotemark{b} \\
60221.993 & HST & STIS & +9.3 \\
60222.376 & SOAR & Goodman & +9.6 \\
60222.480 & MMT & Blue Channel & +9.7 \\
60222.488 & MMT & Blue Channel & +9.7 & \tablenotemark{b} \\
60222.583 & FTN & FLOYDS & +9.8 \\
60223.592 & FTN & FLOYDS & +10.8 \\
60226.571 & FTN & FLOYDS & +13.7 \\
60228.988 & HST & STIS & +16.1 \\
60229.522 & Shane & Kast & +16.6 \\
60230.579 & FTN & FLOYDS & +17.6 \\
60230.615 & Keck I & LRIS & +17.7 \\
60234.539 & Shane & Kast & +21.5 \\
60234.577 & FTN & FLOYDS & +21.5 \\
60236.478 & MMT & Binospec & +23.4 \\
60236.630 & Keck I & LRIS & +23.5 \\
60236.917 & HCT & HFOSC & +23.8 \\
60238.485 & Shane & Kast & +25.3 \\
60239.194 & DOT & ADFOSC & +26.0 \\
60241.194 & DOT & ADFOSC & +28.0 \\
60242.635 & Gemini-N & GNIRS & +29.4 \\
60244.924 & HCT & HFOSC & +31.6 \\
60245.948 & HCT & HFOSC & +32.6 \\
60252.306 & SOAR & Goodman & +38.8 \\
60257.479 & LBT & MODS & +43.8 \\
60258.489 & Shane & Kast & +44.8 \\
60263.456 & MMT & Binospec & +49.6 \\
60266.571 & Keck I & LRIS & +52.7 \\
60290.532 & Keck I & LRIS & +76.0 \\
60292.369 & MMT & Binospec & +77.8 \\
\enddata
\tablenotetext{a}{From \cite{pursiainen_goto_2023}}
\tablenotetext{b}{Higher resolution}
\end{deluxetable}

Two of our spectra obtained with the MMT Blue Channel spectrograph have higher
resolution than the others, with $\Delta\lambda \approx 1.45$\,\AA\ when using
the 1200~lines~mm$^{-1}$ grating. 
From the peaks of the narrow helium P-Cygni lines at 587.6~nm and 667.8~nm in these spectra, we derive a redshift of $z=0.0266$ for SN\,2023tsz. Adopting a flat $\Lambda$CDM cosmology with $\Omega_\mathrm{m,0} = 0.3$ and $H_0 = 70\ \mathrm{km\ s^{-1}\ Mpc^{-1}}$, we use this redshift to derive a distance modulus of $\mu=35.3$~mag, or a luminosity distance of $d_L = 116$~Mpc.

We do not observe any absorption from the \ion{Na}{1}~D doublet at this redshift in the higher-resolution spectra, 
so we neglect extinction from the host galaxy \citep{poznanski_empirical_2012}. We adopt only the Milky Way extinction value of $E(B-V) = 0.0365$~mag \citep{schlafly_measuring_2011}, for which we correct our photometry using a \cite{fitzpatrick_correcting_1999} extinction law.


SN~2023tsz is nearly coincident (angular distance $0\farcs5$) with a faint ($g = 22.8 \pm 0.1$~mag, $r = 22.4 \pm 0.1$~mag) galaxy, PSO~J129.3730-00.0431 \citep{flewelling_pan-starrs1_2020}, which we assume to be the host. The SN must not be associated with the much brighter ($g = 16.2$~mag) galaxy $32\farcs{}5$ to the east because of its much higher redshift ($z=0.035$; \citealt{jones_6df_2009}). We estimate the metallicity of the host galaxy using the luminosity--metallicity relationship of \cite{kirby_universal_2013}. First we convert the PS1 $g$ and $r$ magnitudes to $V = 22.5 \pm 0.1$~mag using Eq.~6 of \cite{tonry_pan-starrs1_2012}. This implies a luminosity of $\log(L / {\rm L}_\sun) = 7.02 \pm 0.04$, using $M_{V,\sun} = 4.84$ mag as the absolute $V$ magnitude of the Sun. Applying Eq.~3 of \cite{kirby_universal_2013}, we find $[\mathrm{Fe}/\mathrm{H}] = -1.4 \pm 0.2$ or $Z = (0.04 \pm 0.02) {\rm Z}_\sun$, similar to some Milky Way and Local Group dwarfs. This value is consistent with the metallicity estimated by \cite{warwick_sn_2025} of $Z = 0.03 {\rm Z}_\sun$. \\

\begin{figure*}
    \centering
    \includegraphics[width=0.99\textwidth]{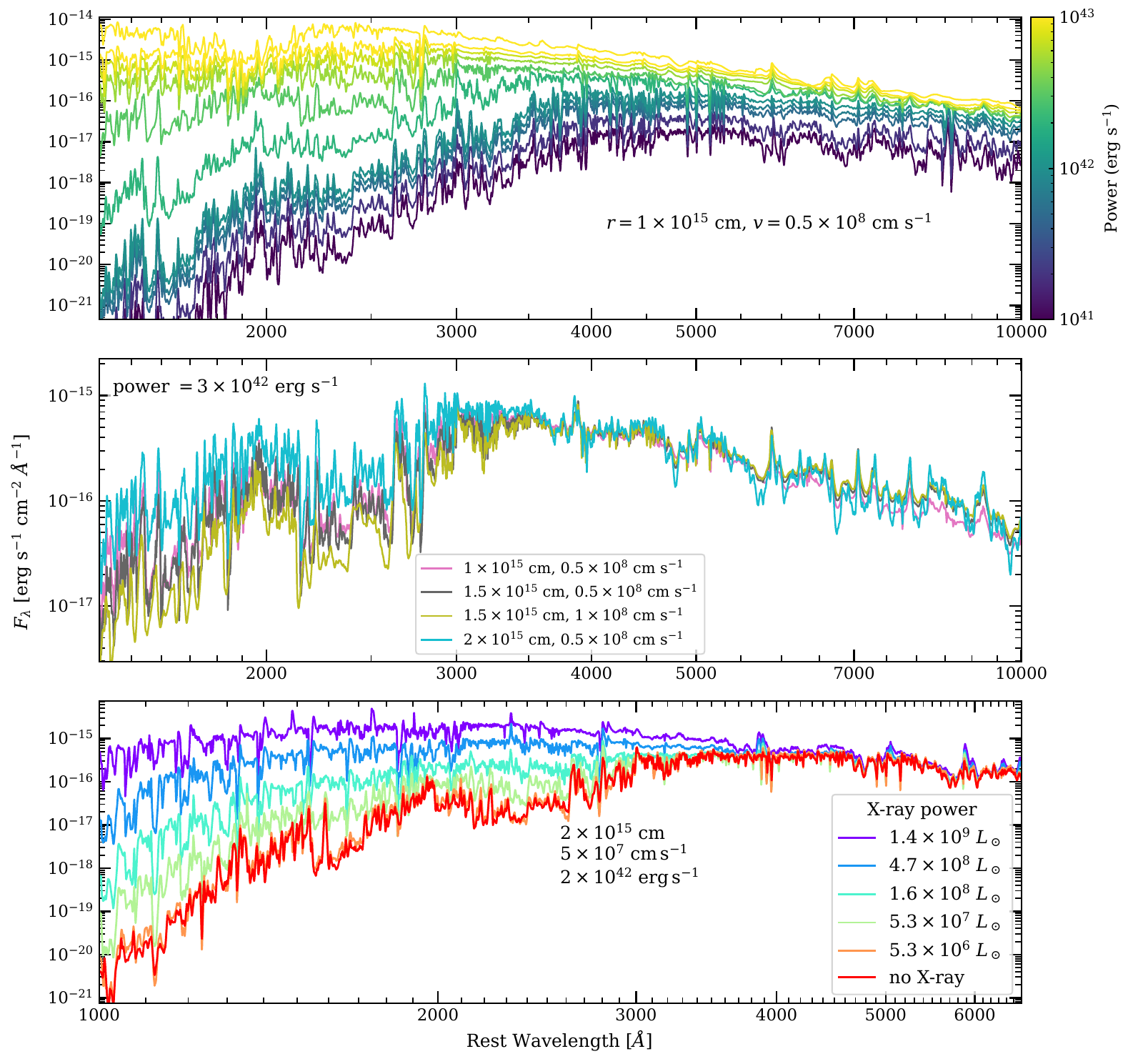}
    \caption{Spectral montage from the model grid based on the exploding 4\,M$_{\odot}$ helium-star model from \citet{dessart_helium_2022} placed at the distance to SN\,2023tsz (116\,Mpc). {\it Top:} Varying power for a fixed radius and velocity, with mixing (See discussion in Section \ref{sec:modeling}). {\it Middle:} Same as top, but instead varying radius and velocity for a fixed power. {\it Bottom:} Comparison of rest-frame UV-optical model spectra for different X-ray irradiation powers.  
    All models share the same radius $r = 2\times10^{15}\,\mathrm{cm}$, CDS velocity $v = 5\times10^{7}\,\mathrm{cm\,s^{-1}}$, and interaction power $ 2\times10^{42}\,\mathrm{erg\,s^{-1}}$; only the imposed X-ray power is varied from the ``no X-ray'' case (red) to the models with a nonzero X-ray field. The X-ray powers are given in units of L$_{\odot}$. The total luminosity is a combination of both the shock power injected in the form of high-energy electrons and the X-ray irradiation field.}
    \label{fig:models}
\end{figure*}

\section{Radiative-Transfer Modeling}
\label{sec:modeling}
We compute synthetic UV–optical-NIR spectra with \textsc{cmfgen} \citep{hillier_treatment_1998,hillier_time-dependent_2012,hillier_photometric_2019,dessart_type_2013}, a one-dimensional (1D), spherically symmetric, non-local-thermodynamic-equilibrium (non-LTE) radiative-transfer code that includes line blanketing and electron scattering. The framework of \citet{dessart_helium_2022} is adopted, in which fast SN ejecta collide with slow-moving CSM, forming a cold dense shell (CDS). The interaction between the ejecta and CSM produces a rapid rise to peak luminosity, while intermediate-width emission lines persist in the spectrum. Such models have been shown to generally match the behavior of some SNe~Ibn \citep{wang_low-mass_2024}.

The analysis in this paper is carried out by comparing the UV-to-NIR observations of SN\,2023tsz to a grid of models based on the \texttt{he4p0} helium-star model used in this framework. This model corresponds to a 4\,M$_{\odot}$ helium star at the start of helium burning \citep{ertl_explosion_2020}, with a pre-SN mass of 3.15\,M$_{\odot}$. The baseline \texttt{he4p0} model has a solar-metallicity composition. 

The choice of this model is motivated by the helium content expected for stripped helium-star progenitors. \citet{dessart_helium_2022} show that the largest fractional helium abundances at core collapse occur for lower-mass binary helium stars, with initial helium-star masses of $\sim2.6$--$5.0\,{\rm M}_{\odot}$. Above this range, more massive helium stars ($\gtrsim5$--$6\,{\rm M}_{\odot}$) have substantially lower fractional helium abundances, as their winds become increasingly dominated by heavier elements, and would therefore not reproduce the strength of the \ion{He}{1} lines relative to the metal features. The lower bound reflects the low-mass end of the helium-star core-collapse models considered \citep[$\gtrsim2.5\,{\rm M}_{\odot}$;][]{woosley_evolution_2019}, below which stripped helium stars fail to produce iron core-collapse explosions. These considerations favor an initial helium-star mass of $\sim2.6$--$5.0\,{\rm M}_{\odot}$, motivating the 4\,M$_{\odot}$ \texttt{he4p0} model adopted here.
A montage of the model spectra is shown in Figure~\ref{fig:models}.

To simulate the interaction luminosity without solving full 3D hydrodynamics, we approximate the shock power as an internal injection of high-energy electrons --- a technique analogous to the treatment of radioactive decay in SN ejecta. This allows us to bypass the geometric complexities of the shock front while retaining a non-LTE treatment of the gas, which is essential for reproducing the optically thin, \ion{Fe}{2}-dominated spectra of SNe Ibn. The models presented in this paper combine those previously analyzed by \citet{dessart_helium_2022} and \citet{wang_low-mass_2024} with new models generated for SN\,2023tsz. 

An important feature of these new models is the inclusion of an X-ray irradiation field ($L_\mathrm{X}$) that is added separately from the prescribed shock power. This provides a more accurate treatment of the microphysics in the CDS. In particular, X-rays act as a source of photoionization,
whereas depositing high-energy electrons leads to nonthermal excitation and ionization. Some of the X-rays escape without being reprocessed, such that the total luminosity is then a combination of the two components: $L_\mathrm{bol} \approx L_\mathrm{sh} + \epsilon L_\mathrm{X}$, where $\epsilon$ is an efficiency parameter close to but less than unity. A montage of spectra with varying X-ray power (in solar luminosities L$_{\odot}$) is shown in the bottom panel of Figure \ref{fig:models}. A presentation of the method is given in Appendix A of \cite{dessart_long-term_2024}.

The model grid further constrains the physical conditions in the interaction region by exploring the effect of macroscopic chemical mixing in the CDS. In particular, comparisons between a mildly mixed CDS and a fully homogeneous composition show that efficient mixing can raise the iron abundance throughout the CDS to about ten times the solar value, which strengthens \ion{Fe}{2} emission in the optical (4000--5500\,\AA) while increasing UV line blanketing (see Figure 11 and Section 4.4 of \citealt{dessart_helium_2022} and Figure \ref{fig:models} of this work). The homogeneous case is interpreted as an upper limit on the degree of mixing. Models with insufficient mixing provide systematically poorer matches to SN~Ibn spectra.

\section{Spectral Analysis and Model Comparison}
\label{sec:results}
\begin{figure*}
    \centering
    \includegraphics[width=0.85\textwidth]{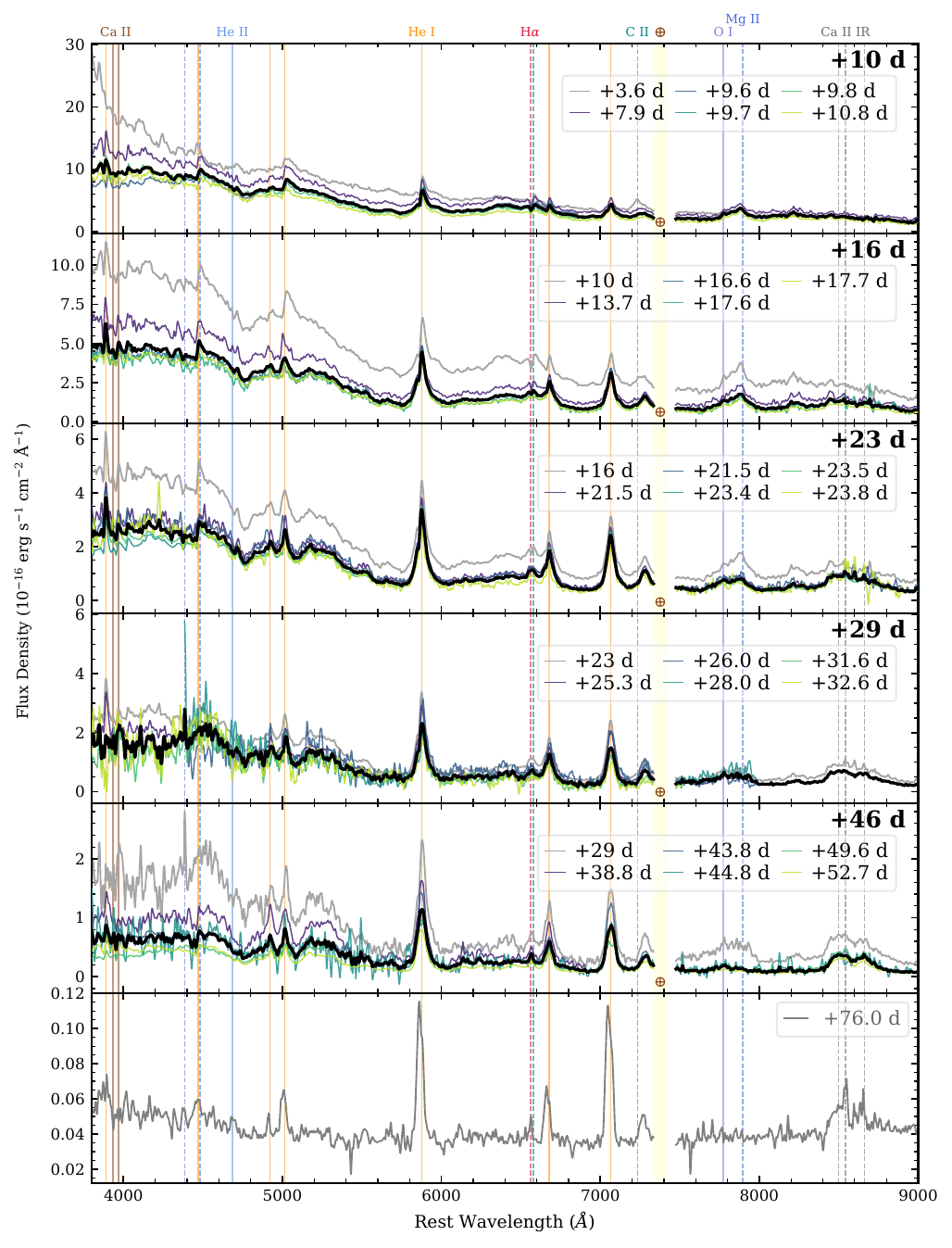}
    \caption{Smoothed, resampled optical spectra, grouped into panels by phase, with major emission lines marked with vertical lines. The $\oplus$ symbol marks wavelengths excised to avoid telluric contamination. In each panel, the latest spectrum from the previous panel is repeated in gray for easy comparison. The remaining spectra (colored lines) are averaged to produce a mean spectrum in each phase bin, plotted in black. Despite the rapidly fading light curve, the identities and intensity ratios of emission features evolve slowly relative to SNe lacking a dense CDS.}
    \label{fig:spec_groups}
\end{figure*}

We select only our low-resolution, high signal-to-noise-ratio (S/N) optical spectra to compare to the model grid. First, we recalibrate each spectrum by multiplying by a constant to match the optical photometry at that phase. Then we smooth the spectrum by convolving with a Gaussian with $\sigma = (0.4~\mathrm{nm}) dx/d\lambda$, where $dx/d\lambda$ is the dispersion function for the spectrograph. For LRIS and MODS, which have two arms with different linear dispersion functions, we smooth each arm separately and then combine the results. Finally, to simplify model comparison, we resample the spectra to a common set of wavelengths: 3800--9000~\AA\ at 2~\AA\ intervals, with a gap in the range 7390--7470~\AA\ to exclude telluric contamination. These smoothed and resampled spectra are plotted in Figure~\ref{fig:spec_groups}, grouped into panels by phase. For the model comparisons, the observed spectra are dereddened and presented in the rest frame.

\begin{figure*}
    \centering
    \includegraphics[width=0.85\textwidth]{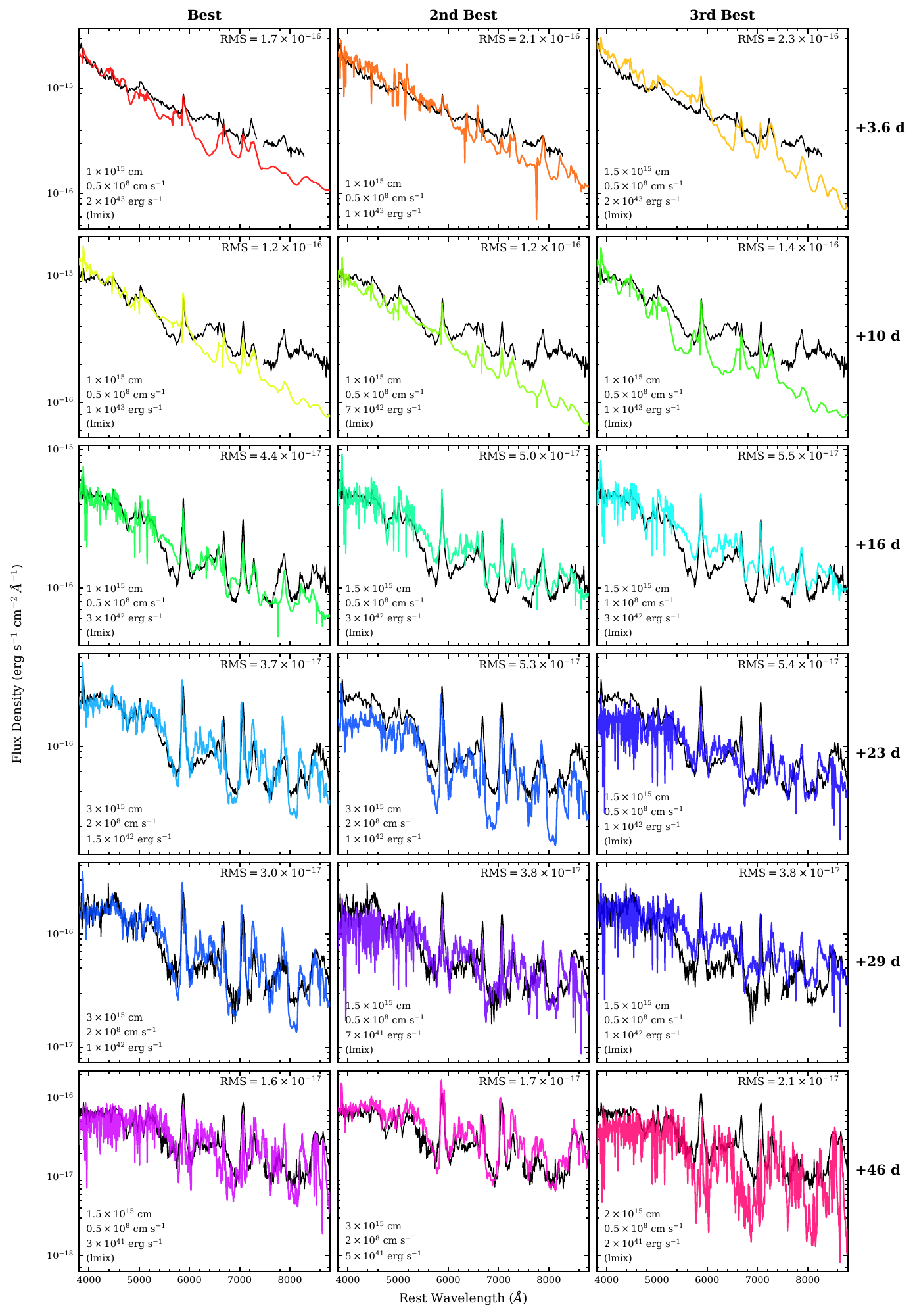}
    \caption{The three models with the lowest $\mathrm{RMS}$ score for each phase, ordered best to third-best from left to right, with the phase shown at the right of each row. For each panel, $\mathrm{RMS}$ is the root-mean-square flux residual of that epoch; the models are scaled to a distance of 116 Mpc with no free normalization. Observed spectra (black lines) are corrected for extinction and displayed in the rest frame. Each panel shows the radius, velocity, and power of the model (``lmix'' indicates a mildly mixed CDS)  in the bottom left, and the $\mathrm{RMS}$ of the fit in the top right. The $\mathrm{RMS}$ score does not fully capture the goodness of fit; for example, we prefer the second- and third-best fits to our earliest spectrum owing to the lack of very narrow lines.}
    \label{fig:best_fit}
\end{figure*}

\begin{figure}
    \centering
    \includegraphics[width=\columnwidth]{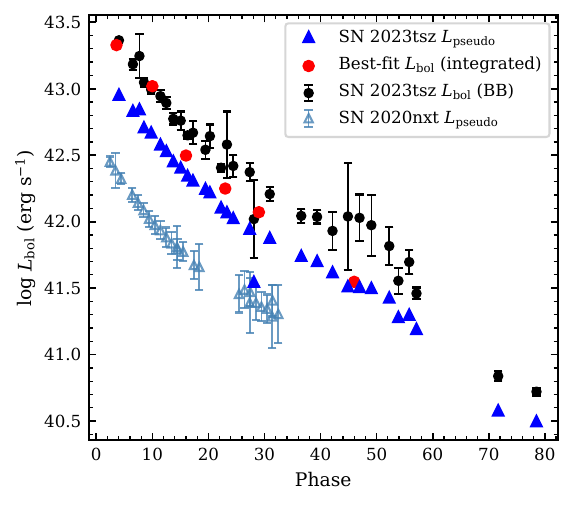}
    \caption{Bolometric (black) and pseudobolometric (filled blue) light curves of SN\,2023tsz, computed with the Light Curve Fitting package \citep{hosseinzadeh_light_2024}, with SN\,2020nxt (open blue) for comparison from \citet{wang_low-mass_2024}. Red points show the integrated total emergent luminosity of the best-fit models from Figure \ref{fig:best_fit}. Error bars are $1\sigma$.}
    \label{fig:best_fit_lbol}
\end{figure}

One striking feature of Figure~\ref{fig:spec_groups} is the relatively slow evolution of the emission lines over time, in stark contrast to the rapidly fading light curve in Figure~\ref{fig:phot}. Motivated by the similarity of the spectra within each panel, we average our data to obtain high-S/N mean spectra at +10, +16, +23, +29, and +46 days after maximum light, to which we add the individual classification spectrum at +3.6 days after peak. 
We compare these spectra to the model grid by calculating a root-mean-squared score $\mathrm{RMS}$, with each model placed at the luminosity distance of SN\,2023tsz ($d = 116$\,Mpc). 
The results of the fit are shown in Figure \ref{fig:best_fit}.

For earlier epochs, the models with a greater interaction power and smaller radii yield a better fit to the data. For later epochs, a lower interaction power and greater radius is needed to fit the observed spectrum. This trend reflects the expansion of the CDS and was similarly shown for SN\,2006jc by \citet{dessart_helium_2022}.  The poorer agreement at the earliest epochs may reflect the fact that these
phases are not well described by the simplified CDS configuration adopted in
the model grid. 

To compare the energetics of our best-fit models with the observed luminosity, we construct the bolometric and pseudobolometric light curves of SN\,2023tsz with the Light Curve Fitting package \citep{hosseinzadeh_light_2024}, which fits a blackbody to the extinction-corrected broadband spectral energy distribution (SED) at each epoch. The full bolometric luminosity is the Stefan--Boltzmann luminosity of the best-fit blackbody, while the pseudobolometric luminosity is obtained by integrating the best-fit blackbody between the $U$ and $I$ bands.
Figure~\ref{fig:best_fit_lbol} compares these light curves to those of the Type~Ibn SN\,2020nxt \citep{wang_low-mass_2024} and to the luminosities of our best-fit models from Figure~\ref{fig:best_fit}. For each model we plot the total emergent luminosity, obtained by integrating the model SED.
The best-fit model luminosities are broadly consistent with the observed bolometric light curve.
\subsection{The Near-Infrared Spectrum}

\begin{figure*}
    \centering
    \includegraphics[width=0.95\textwidth]{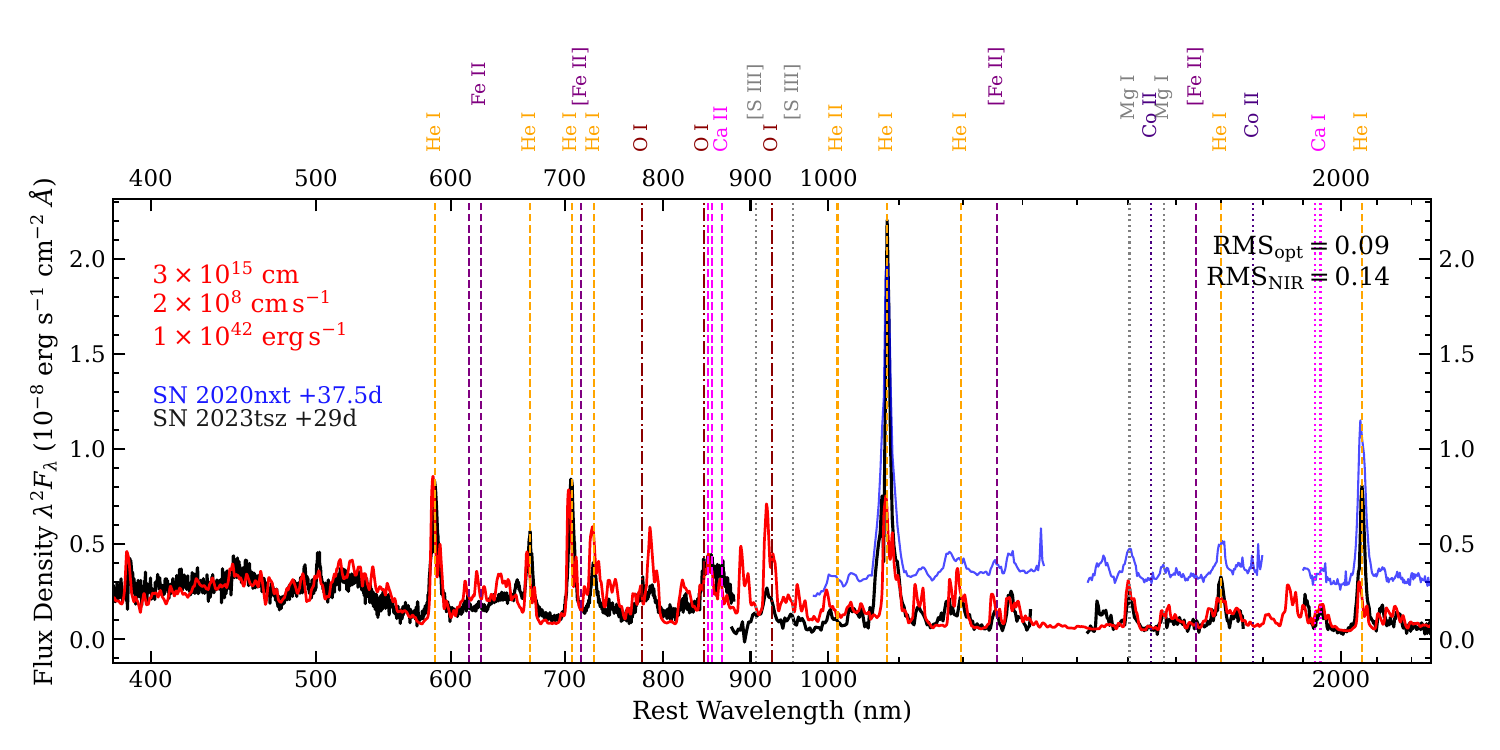}
    \caption{The complete optical-to-NIR spectrum of SN\,2023tsz at +29\,d (black) compared to a model (red) with $r=3\times10^{15}$~cm, $v=2\times10^8$~cm~s$^{-1}$, and $\mathrm{power}=1\times10^{42}$~erg~s$^{-1}$. This same model is independently the best fit to both the optical and NIR portions of the spectrum at this phase. A Keck NIRES spectrum of SN\,2020nxt on day +37.5 from \citet{wang_low-mass_2024} is shown in blue. Both spectra exhibit a number of intermediate-width P-Cygni profiles of He I (marked as the vertical dashed lines in yellow), oxygen lines (as dashed-dotted lines in brown color), and other lines such as Mg I, Fe II, Cal I, and C I.}
    \label{fig:best_fit_nir_new}
\end{figure*}
The NIR spectrum of Type Ibn SN\,2023tsz is shown in Figure
\ref{fig:best_fit_nir_new}. It exhibits many He I lines, including
strong He I features at 1.0830, 1.869, and 2.0581 $\mu$m, as well as the lines at
 1.197, 1.278, 1.508, 1.7, 1.909, and 2.112 $\mu$m. The
spectrum also shows O I lines at 1.13 $\mu$m and other features due to Mg I, Fe
II, Ca I, and C I. We compare the spectrum of the Type Ibn SN\,2023tsz with that
of the Type Ibn SN\,2020nxt \citep{wang_low-mass_2024} highlighting the similarity between these two events.

We also independently fit the optical and NIR spectra at +29 d, finding remarkable agreement: the best-fit model is the same for both datasets, corresponding to $r_{\rm CSM}=2\times10^{15}$ cm, $v=2\times10^{8}$ cm s$^{-1}$, and a shock power of $10^{42}$ erg s$^{-1}$ (see Figure \ref{fig:best_fit_nir_new}). In particular, the model matches the relative strengths of many \ion{He}{1} lines and \ion{Mg}{1}. Given that a relatively simple model construction was used (i.e., without complex 3D hydrodynamics/mixing, which may become important at this phase), this level of agreement is striking.

\subsection{The UV Spectrum}
\begin{figure*}
    \centering
    \includegraphics[width=\textwidth]{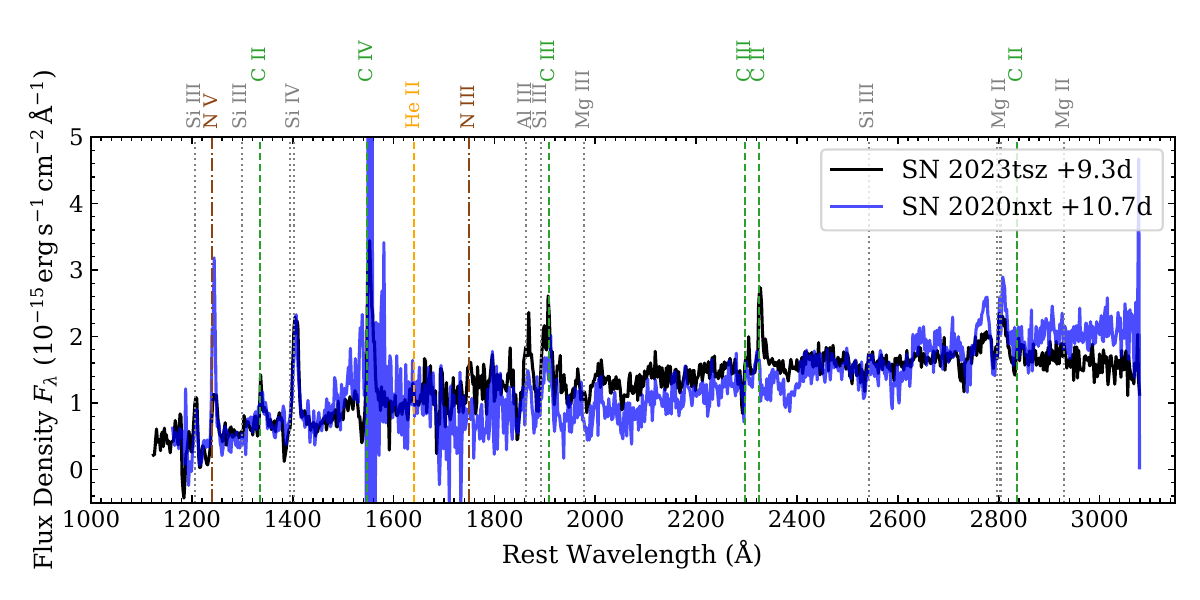}
    \caption{UV spectra with lines labeled. HST/STIS  total flux density spectra of SN\,2023tsz (black) in the far-UV regions, 
    corresponding to +9.3\,d after maximum light. The spectrum is compared to that  of SN\,2020nxt (blue) at +10.7 days, which was scaled by a factor of 1.7 to match the continuum \citep{wang_low-mass_2024}. Both spectra show very similar features and line velocities, suggesting that the bulk of the radiation arises from material with similar kinematics. This is especially evident in the similarly shaped carbon features, such as \ion{C}{2}/\ion{}{3}/\ion{}{4}, as well as the \ion{Si}{4} and \ion{Mg}{2} lines.}
    \label{fig:uvspec_20nxt}
\end{figure*}

\begin{figure}
    \centering
    \includegraphics[width=0.5\textwidth]{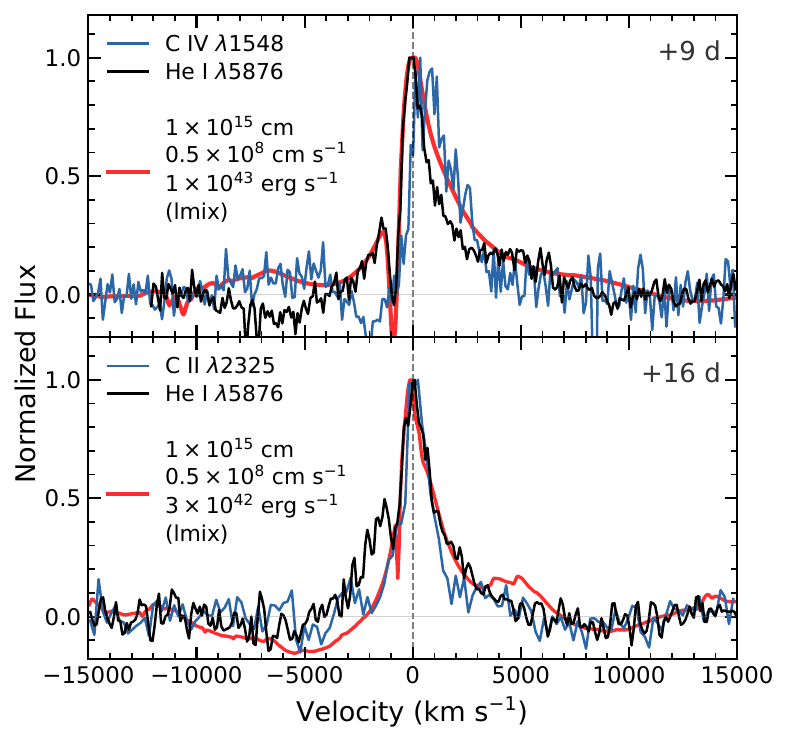}
    \caption{ Comparison of continuum-normalized line profiles for SN\,2023tsz in
velocity space, grouped by epoch. The spectra are shown in the rest frame,
centered on each line. \textit{Top:} C\,IV~$\lambda1548$ (blue) and
He\,I~$\lambda5876$ (black) at $+9.3$~d. \textit{Bottom:} C\,II~$\lambda2325$
(blue) and He\,I~$\lambda5876$ (black) at $+16.1$~d (relative to maximum light).
 The red curve in each panel shows the He\,I~$\lambda5876$ profile from the best-fitting model at the nearest computed epoch, with the model CSM radius, velocity, and injected power annotated. The similar widths of the UV carbon and optical helium lines are
consistent with a common formation region in the CDS. }
    \label{fig:uvo_comp}
\end{figure}
Our UV observations of SN\,2023tsz on days +9.3 and +16.1 after maximum light reveal spectra that are markedly different from what is seen in weak to noninteracting hydrogen-rich SNe \citep{vasylyev_early-time_2022,vasylyev_early-time_2023,bostroem_sn_2023} at comparable epochs. The FUV and NUV spectra are dominated by a smooth continuum with intermediate-width P-Cygni line profiles, most prominently from highly ionized species such as \ion{C}{3}, \ion{C}{4}, \ion{Si}{3}, \ion{Si}{4}, \ion{Al}{3}, and \ion{N}{5}, along with lower-ionization \ion{Mg}{2} $\lambda2798$ and \ion{C}{2} features also dominating the spectrum.

Observations of the Type Ibn SN\,2020nxt also showed strong lines of highly ionized species \citep{wang_low-mass_2024}, including the \ion{C}{4} $\lambda\lambda$1548, 1550 doublet, \ion{N}{5} $\lambda$1240, and features likely associated with \ion{Si}{4} or \ion{O}{4} around 1400\,\AA\ (Figure \ref{fig:uvspec_20nxt}). The similarity between SN\,2023tsz and SN\,2020nxt suggests they had a similar progenitor. These UV lines have FWHM velocities ($\sim 2500$--$3000$\,km\,s$^{-1}$) that are comparable to the widths of the \ion{He}{1} emission lines observed in the optical (Figure \ref{fig:uvo_comp}). This similarity indicates that both the high-ionization FUV lines and the optical \ion{He}{1} lines share a common formation region within the CDS, rather than originating from unshocked, distinct wind components.

The persistence of strong UV flux and intermediate-width UV line profiles in SN\,2023tsz therefore points to an interaction-powered origin rather than simple cooling-envelope emission. More generally, the UV provides a useful discriminant between pure SN shock cooling and scenarios powered by ejecta--CSM interaction. In weakly or noninteracting SNe~II, such as SN\,2022acko \citep{bostroem_early_2023} and SN\,2022wsp \citep{vasylyev_early-time_2023}, the early UV evolution within the first two weeks is characterized by a rapid decline in flux and the presence of broad P-Cygni line profiles. As the ejecta cool, a ``forest'' of metal absorption lines, primarily from iron-group elements, develops, resulting in strong line blanketing that suppresses the flux shortward of 3000\,\AA. However, some objects like SN\,2021yja \citep{vasylyev_early-time_2022} show a significant UV excess, hinting at interaction with CSM within one week after explosion.

In contrast, interaction with dense CSM can sustain the UV luminosity significantly longer than in cooling-envelope scenarios \citep{dessart_modeling_2022}. This effect was clearly observed in SN\,2023ixf, where the interaction between the ejecta and confined CSM produced strong UV line emission that persisted well beyond the timescales typical for noninteracting events, such as the development of a boxy \ion{Mg}{2} emission feature \citep{bostroem_circumstellar_2024}.

\begin{figure*}
    \centering
    \includegraphics[width=0.9\textwidth]{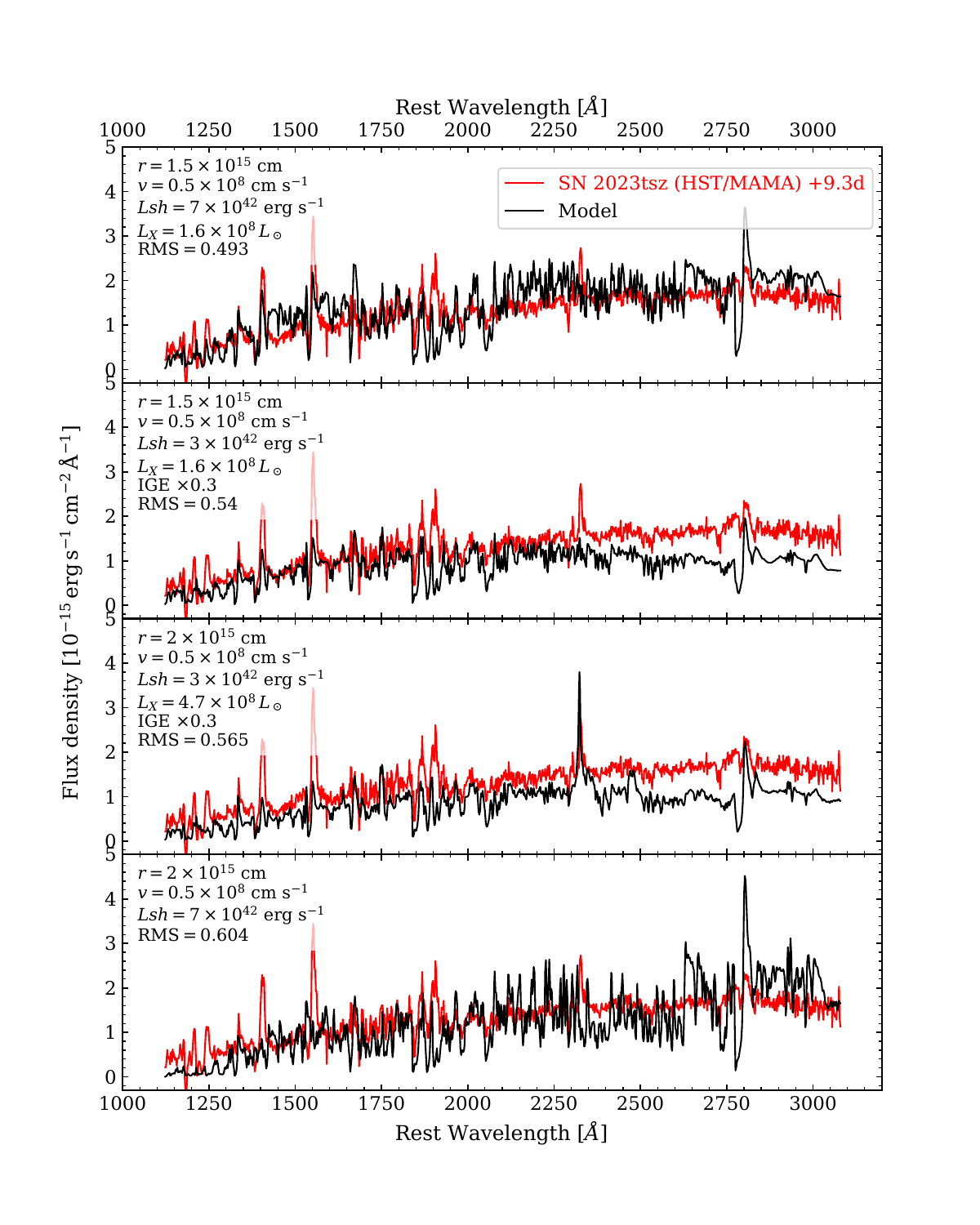}
    \caption{Best-fitting model spectra (black) for observed FUV+NUV HST/MAMA spectrum of SN\,2023tsz at +9.3 days (red) after maximum light. The best-fit model corresponds to an exploding helium star (4\,M$_{\odot}$ initial mass at He burning) with radius $r=2\times10^{15}$~cm, velocity $v=5\times10^7$~cm~s$^{-1}$, $\mathrm{power}=1\times10^{42}$~erg~s$^{-1}$, X-ray power $1.6\times10^{8}\,{\rm L}_{\odot}$, and iron-group element abundance 
    scaled by 0.3. The panels are ordered by lowest (top) to highest (bottom) $\mathrm{RMS}$ residuals.}
    \label{fig:best_fit_uv1}
\end{figure*}

\begin{figure*}
    \centering
    \includegraphics[width=0.9\textwidth]{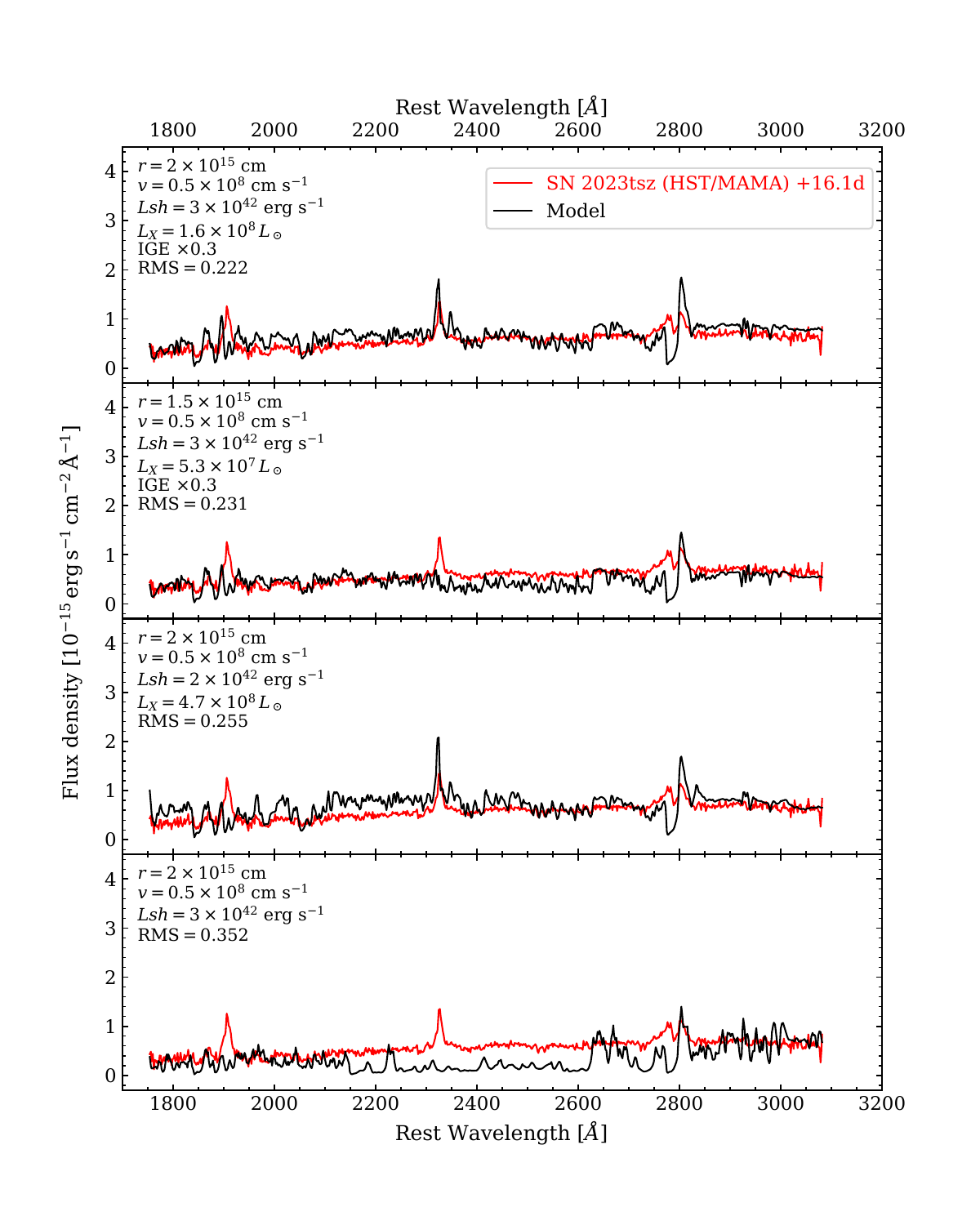}
    \caption{Same as Figure \ref{fig:best_fit_uv1} but for +16.1 days after maximum.}
    \label{fig:best_fit_uv2}
\end{figure*}

\subsection{Line-Profile Evolution and the Cold Dense Shell}
There are important subtleties in comparing the intermediate-width line profiles observed in SN\,2023tsz, and in other Type Ibn SNe, to the Lorentzian electron-scattering profiles often observed in hydrogen-rich interacting SNe. SNe~II often show emission-line cores, with FWHM values up to an order of magnitude smaller than those discussed here \citep{jacobson-galan_final_2024}. At early times, these narrow recombination lines are frequently accompanied by broader, Lorentzian-like wings attributed to electron scattering of line photons in dense CSM. These features typically evolve on relatively short timescales, broadening and fading as the CSM is swept up. However, the Lorentzian narrow lines have also been observed in some SNe~II, lasting up to months after explosion (SNe~IIn). The intermediate-width lines observed in SN\,2023tsz, and in other Type Ibn SNe, show remarkably little change in width or relative strength over the course of two months in the optical. In SNe Ibn, electron-scattering broadening should contribute less to the line flux than in hydrogen-rich SNe IIn because helium-rich gas provides fewer free electrons per unit mass, and because the line-forming region is the fast-moving CDS rather than a slow, extended, unshocked CSM. 
The lower ionization, in particular at later times, also inhibits
electron-scattering broadening. 
The models of \citet{dessart_helium_2022}, together with those presented here for SN\,2023tsz, support a paradigm in which a sufficiently massive outer shell/CSM is needed to decelerate the ejecta, causing most of the mass to pile up into a CDS and thereby preventing the emergence of broad late-time line features.

\begin{figure*}
    \centering
    \includegraphics[width=0.9\textwidth]{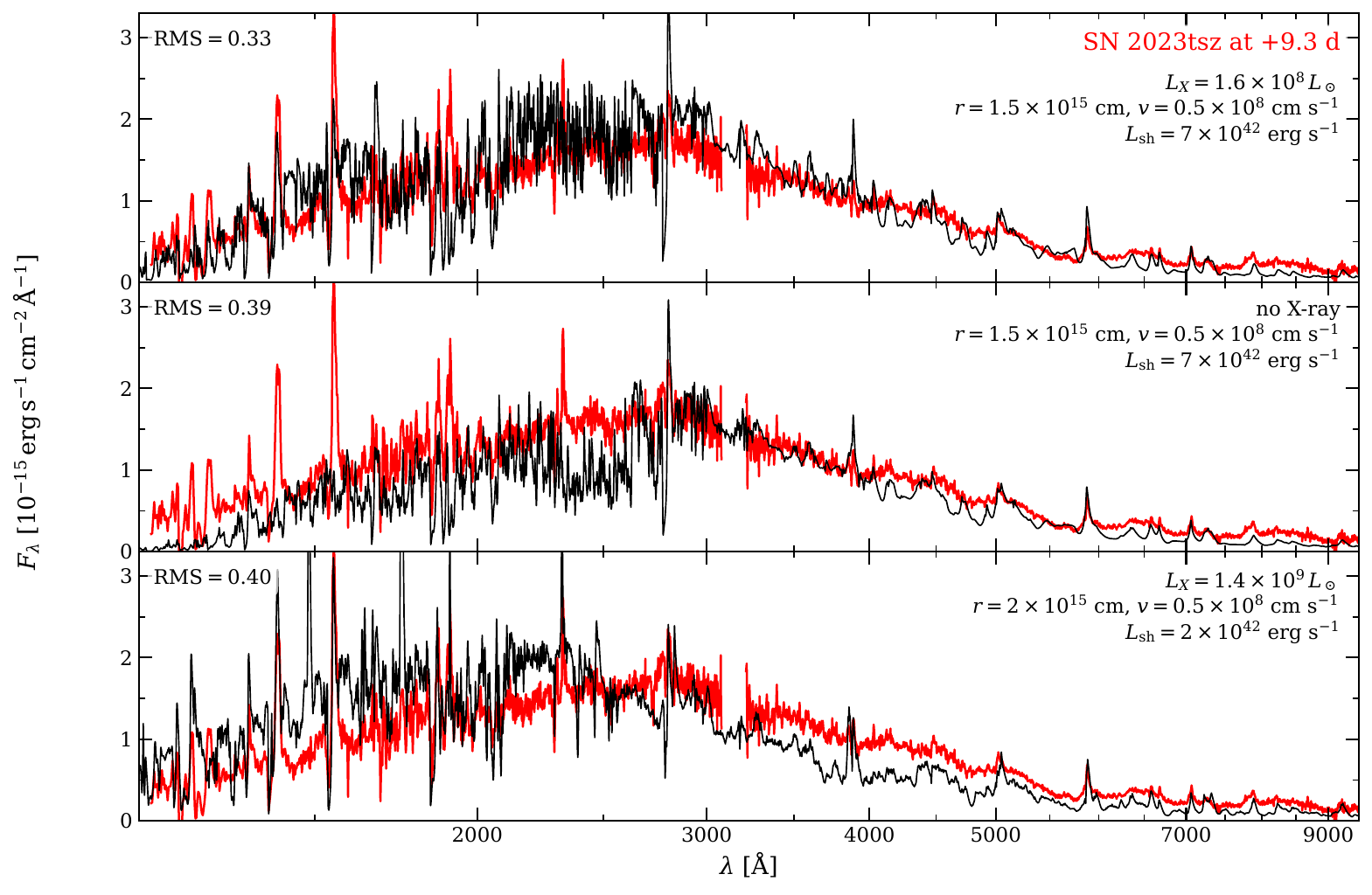}
    \caption{Top-three best-fitting model spectra of 23tsz (black) to the observed STIS/HST MAMA UV spectrum on day 9.3 stitched together with the optical (red). The UV and optical were fit simultaneously.}
    \label{fig:best_fit_uvo}
\end{figure*}

For SNe Ibn, the SED below $\sim5500$\,\AA\ is highly sensitive to the iron-group content of the line-forming CDS. As demonstrated by \citet{dessart_helium_2022}, the characteristic pseudocontinuum in this region is largely produced by a forest of \ion{Fe}{2} transitions, and models with very low metal content do not reproduce the observed \ion{Fe}{2}-dominated emission. Thus, the strong blue/NUV metal-line emission in SN\,2023tsz argues against an extremely metal-poor line-forming region. However, this should not be interpreted as requiring a solar-metallicity progenitor. Some of our preferred UV models use an iron-group abundance scaling of 0.3 relative to the nominal (solar metallicity) model, indicating that moderately reduced iron-group abundances can still reproduce the data.
This also emphasizes the unnecessary need for pollution
by explosively produced metals like Ni, Co, or Fe, either in a stable
or unstable form -- a uniform mass fraction of Fe of a few $\times 10^4$ is sufficient
to produce this large \ion{Fe}{2} emission bump in SNe~Ibn a few weeks past peak.

\subsection{Fitting Models to the UV Spectra}
We fit the UV spectra on days +9.3 and +16 using the same approach as for the optical-NIR. The results of the fit are shown in Figures \ref{fig:best_fit_uv1} and \ref{fig:best_fit_uv2}, respectively. Reproducing the UV continuum and line features is challenging, particularly in models requiring the simultaneous presence of \ion{C}{2}, \ion{C}{3}, and \ion{C} {4}. In this regime, the ionization balance is extremely sensitive to the incident X-ray flux; minor fluctuations can cause these UV lines to appear or vanish while leaving the optical spectrum largely unaffected. Thus, the models can be tuned such that they fit the relative strength of the emission lines and the continuum in the UV and optical-NIR. The UV spectra are best reproduced by models with $L_\mathrm{X}\approx 10^{8}\,{\rm L}_{\odot}$, specifically $1.6\times10^{8}\,{\rm L}_{\odot}$ and  $4.7\times10^{8}\,{\rm L}_{\odot}$.

Figure \ref{fig:best_fit_uvo} shows the top-three best-fitting models to the FUV, NUV, and optical data simultaneously on day +9.3. The top-two panels show models with a CDS radius of $r_{\rm csm} = 1.5 \times 10^{15}$\,cm and an interaction power of $7 \times 10^{42}$\,erg\,s$^{-1}$ ($L_{\rm sh}$). The only difference between these two models is that the upper model includes an X-ray irradiation field of $L_\mathrm{X} = 1.6\times10^{8}\,{\rm L}_{\odot}$, while the lower model does not include X-rays.
The third-best model has a larger  CDS radius of $2 \times 10^{15}$\,cm, a lower X-ray contribution of  $1.4\times10^{9}\,{\rm L}_{\odot}$, but at a lower interaction power of $2 \times 10 ^{42}$ ergs s$^{-1}$.  



\section{Conclusions}
\label{sec:conclusions}
We present UV, optical, and NIR spectra of the nearby Type~Ibn SN\,2023tsz, including two epochs of HST/STIS UV within the first 2 weeks after maximum light. These data are complemented with extensive ground-based optical-NIR follow-up observations to $\sim 3$ months after first detection. The optical-NIR spectra exhibit intermediate-width \ion{He}{1} emission features, indicating that the observed radiation is dominated by interaction with dense, hydrogen-poor but helium-rich CSM. Despite the rapidly fading light curve, the intermediate-width emission lines evolve slowly over the first several weeks, with similar line-forming conditions over this time, particularly in a CDS. This physical picture is consistent with He-star explosion models that favor the formation of such a CDS to keep the intermediate-width lines evolving slowly, prevent the emergence of broad P-Cygni features, and reproduce the simultaneous presence of \ion{C}{2} and \ion{C}{4} in the UV at early times.

We compared the spectra to a grid of 1D non-LTE \textsc{cmfgen} interaction models in which fast SN ejecta collide with a dense, helium-rich CSM shell. The observed continua and the majority of the prominent emission features are well reproduced within this framework. The preferred models shift from higher interaction power and smaller radii at early times to lower power and larger radii at later times, mirroring the behavior found for SN\,2006jc by \citet{dessart_helium_2022}.

The interaction luminosity is approximated by internal deposition of high-energy electrons. While this treatment reproduces much of the optical--NIR continuum and line strengths, we find that matching the highly ionized UV features requires a separately added X-ray irradiation field, $L_\mathrm{X}$, which provides photoionization distinct from the nonthermal excitation and ionization produced by electron deposition. With this X-ray component, we find the models better describe the relative line strengths in the UV and optical. Explosion models using a 4\,M$_{\odot}$ He-star initial mass (corresponding to a pre-SN mass 3.15 M$_{\odot}$  and a ZAMS mass 18 M$_{\odot}$) are able to reproduce the strong \ion{He}{1} lines in the optical spectra while also matching the C lines in the UV.  Additional model parameters are tuned, including mixing, the position and velocity of the CDS, and the X-ray irradiation power. These results add to the growing case for lower-mass, binary-stripped helium stars as the progenitors of some SNe Ibn. The low-luminosity, low-metallicity dwarf host galaxy and low ejecta mass \citep{warwick_sn_2025} are more compatible with a binary-stripped lower-mass channel than with a single very massive WR wind-stripping scenario. 

Although our modeling favors a low-mass helium-star progenitor, we caution that the inferred parameters are subject to degeneracies, particularly between the helium-star mass and the properties of the CDS \citep{dessart_helium_2022}. For example, a model with a larger radius and higher CDS mass may be difficult to distinguish from one with a smaller radius and lower CDS mass. The helium-star mass is nonetheless constrained from both ends. As discussed in Section~\ref{sec:modeling}, the strength of the \ion{He}{1} lines relative to the metal features favors the lower-mass helium-star regime, while the low-mass end is bounded by the requirement that the stripped helium star undergo iron core collapse. These considerations favor an initial helium-star mass of $\sim2.6$--$5.0\,{\rm M}_\odot$, motivating the 4\,M$_\odot$ model adopted here. The light curve, though itself highly degenerate, independently points to a relatively low ejecta mass \citep[$\sim0.6\,{\rm M}_{\odot}$;][]{warwick_sn_2025}, indicating a modest explosion and reinforcing a lower-mass progenitor.

\bigskip
\medskip

Time-domain research by the University of Arizona
team and D.J.S. is supported by National Science Foundation
(NSF) grants 2308181, 2407566, and 2432036.
Q.W. is supported by the Sagol Weizmann-MIT Bridge Program.
Supernova research at Rutgers University is supported in part by the NSF award AST-2407567.
Y.Y.'s research is now partially supported by the Tsinghua University Dushi Program, and was previously supported through a Benoziyo Prize Postdoctoral Fellowship and the Bengier-Winslow-Robertson Fellowship. 
A.V.F.’s research group at UC Berkeley acknowledges financial 
assistance from NASA/HST grant GO-17195, as well as from the Christopher R. Redlich Fund, Gary and Cynthia
 Bengier, Clark and Sharon Winslow, Alan Eustace and Kathy Kwan (W.Z. is a Bengier-Winslow-Eustace Specialist in Astronomy), Timothy and Melissa Draper, Briggs and Kathleen Wood, Ellyn and Alan 
Seelenfreund (T.G.B. is Draper-Wood-Seelenfreund Specialist in 
Astronomy), and numerous other donors. 

We appreciate the expert assistance of the staff at the various observatories where data were obtained.   
This work is based in part on observations made with the NASA/ESA Hubble Space Telescope, obtained from the Data Archive at the Space Telescope Science Institute (STScI), which is operated by the Association of Universities for Research in Astronomy (AURA), Inc., under NASA contract NAS 05-26555.
Some of the data presented herein were obtained at the W. M. Keck
Observatory, which is operated as a scientific partnership among the
California Institute of Technology, the University of California, and
NASA; the observatory was made possible by the generous financial
support of the W. M. Keck Foundation.
A major upgrade of the Kast spectrograph on the Shane 3\,m telescope at Lick Observatory, led by Brad Holden, was made possible through gifts from the Heising-Simons Foundation, 
William and Marina Kast, and the University of California Observatories. 
We thank the following U.C. Berkeley undergraduate students for assistance with some of the Lick/Nickel observations: Emma Born, Elma Chuang, Cooper Jacobus, Sophia Risin, Yoomee Zeng.
Research at Lick Observatory is partially supported by a gift from Google.

\newpage
\appendix


\section{Photometric Reduction} 
\label{app:phot}
Photometry from the Las Cumbres Observatory (LCO) was reduced using \texttt{lcogtsnpipe}, a PyRAF-based photometric reduction pipeline \citep{valenti_diversity_2016}. Point-spread-function (PSF) photometry was performed, and the $UBV$ data were calibrated to Vega magnitudes using standard fields observed on the same night with the same telescope \citep{stetson_homogeneous_2000}. KAIT images were reduced using the custom pipeline described by \cite{stahl_lick_2019}. PSF photometry was obtained using DAOPHOT \citep{stetson_daophot_1987} from the IDL Astronomy Users Library. The \cite{landolt_ubvri_1992} magnitudes were transformed to the KAIT/Nickel natural system before calibration. Apparent magnitudes were measured in the KAIT4/Nickel2 natural system and transformed to the standard system using the local calibrator and color terms for KAIT4 and Nickel2 \citep[see][]{stahl_lick_2019}.

For the HOWPol data, PSF photometry was performed as template images were not available. The DFOT observations were reduced using an IRAF- and DAOPHOT II-based routine developed by \cite{roy_sn_2011}. 
The \textit{Swift}/UVOT data were downloaded from the NASA Swift Data Archive\footnote{\url{https://heasarc.gsfc.nasa.gov/cgi-bin/W3Browse/swift.pl}} and reduced using standard software distributed with HEAsoft.\footnote{\url{https://heasarc.gsfc.nasa.gov/docs/software/heasoft/}} 

The NIR data from kSIRIUS were reduced using standard procedures in IRAF. Photometric magnitudes were measured using PSF photometry with standard IRAF tasks, including DAOPHOT \citep{stetson_daophot_1987}, and calibrated using secondary stars from the 2MASS catalog \citep{skrutskie_two_2006}.

\section{Spectroscopic Reduction}
\label{app:spec}
We reduced the LCO/FLOYDS spectra using the \texttt{floydsspec} pipeline,\footnote{\url{https://github.com/svalenti/FLOYDS_pipeline}} which performs standard CCD processing, 1D extraction, wavelength calibration, and flux calibration \citep{valenti_first_2014}.

The Kast spectra were reduced following standard techniques for CCD processing and spectrum extraction \citep{silverman_berkeley_2012} using IRAF \citep{national_optical_astronomy_observatories_iraf_1999}
~routines and custom Python and IDL codes.\footnote{\url{https://github.com/ishivvers/TheKastShiv}} Low-order polynomial fits to comparison-lamp spectra were used to calibrate the wavelength scale, and small adjustments derived from night-sky emission lines in the target frames were applied. The spectra were flux calibrated using observations of spectrophotometric standard stars obtained on the same night, at similar airmass, and with the same instrument configuration. For these observations, the long slit was oriented at or near the parallactic angle to minimize slit losses caused by atmospheric dispersion \citep{filippenko_importance_1982}.

We reduced the MMT/Binospec spectra using the Binospec pipeline \citep{kansky_binospec_2019}. The MMT/Blue Channel spectra were reduced using standard IRAF routines, including bias subtraction, flat-field correction, 1D extraction, wavelength calibration, and flux calibration. We reduced the SOAR/Goodman spectra using the Goodman spectroscopic pipeline \citep{torres_goodman_2017}, which performs standard image processing, spectral extraction, wavelength calibration, and flux calibration. The Keck/LRIS spectra were reduced using \texttt{LPipe} \citep{perley_fully_2019}. 
We reduced the DOT spectra using standard tasks in IRAF, including preprocessing, extraction of the 1D spectra, wavelength calibration, and flux calibration.
The Gemini/GNIRS spectrum was reduced using the Gemini IRAF package. The spectrum was telluric and flux calibrated order by order using \texttt{xtellcor} \citep{vacca_method_2003} in the \texttt{Spextool} package \citep{cushing_spextool_2004}, with an A0\,V standard star observed at a similar time.

\bibliography{merged}

\end{document}